\documentclass[conference,hidelinks]{IEEEtran}

\usepackage{graphicx,url}
\usepackage[english]{babel}
\addto\captionsenglish{}
\usepackage[utf8]{inputenc}
\usepackage{amsmath}
\usepackage{amssymb}
\usepackage{lipsum}
\usepackage{mathtools}
\usepackage{cuted}
\usepackage{cite}
\usepackage{graphicx}
\usepackage{balance}
\usepackage{tikz}
\usepackage{pgfplots}
\usepackage{tikzscale} 
\pgfplotsset{compat=newest} 
\usepackage{scalefnt}
\usepackage{orcidlink}
\usepackage{dblfloatfix}
\usepackage[acronym,shortcuts]{glossaries}

\newacronym{RPE}{RPE}{radar parameter estimation}
\newacronym{OTFS}{OTFS}{orthogonal time frequency space}
\newacronym{AFDM}{AFDM}{affine frequency division multiplexing}
\newacronym{MIMO}{MIMO}{multiple-input multiple-output}
\newacronym{SISO}{SISO}{single-input single-output}
\newacronym{ISAC}{ISAC}{integrated sensing and communications}
\newacronym{3D}{3D}{three-dimensional}
\newacronym{2D}{2D}{two-dimensional}
\newacronym{1D}{1D}{one-dimensional}
\newacronym{RX}{RX}{receiver}
\newacronym{TX}{TX}{transmitter}
\newacronym{BF}{BF}{beamforming}
\newacronym{mmWave}{mmWave}{millimeter-wave}
\newacronym{SotA}{SotA}{state-of-the-art}
\newacronym{ULA}{ULA}{uniform linear array}
\newacronym{QAM}{QAM}{quadrature amplitude modulation}
\newacronym{ISFFT}{ISFFT}{inverse symplectic finite Fourier transform}
\newacronym{SFFT}{SFFT}{symplectic finite Fourier transform}
\newacronym{AWGN}{AWGN}{additive white Gaussian noise}
\newacronym{OFDM}{OFDM}{orthogonal frequency division multiplexing}
\newacronym{OCDM}{OCDM}{orthogonal chirp division multiplexing}
\newacronym{BS}{BS}{base station}
\newacronym{UE}{UE}{user equipment}
\newacronym{DFT}{DFT}{discrete Fourier transform}
\newacronym{IDFT}{IDFT}{inverse discrete Fourier transform}
\newacronym{IFFT}{IFFT}{inverse fast Fourier transform}
\newacronym{TD}{TD}{time-domain}
\newacronym{wlg}{wlg}{without loss of generality}
\newacronym{CP}{CP}{cyclic prefix}
\newacronym{DAFT}{DAFT}{discrete affine Fourier transform}
\newacronym{DAF}{DAF}{discrete affine Fourier}
\newacronym{IDAFT}{IDAFT}{inverse discrete affine Fourier transform}
\newacronym{CPP}{CPP}{\textit{chirp-periodic} prefix}
\newacronym{IDZT}{IDZT}{inverse discrete Zak transform}
\newacronym{DZT}{DZT}{discrete Zak transform}
\newacronym{ICI}{ICI}{inter-carrier interference}
\newacronym{BER}{BER}{bit error rate}
\newacronym{DoF}{DoF}{degrees-of-freedom}
\newacronym{FD}{FD}{full-duplex}
\newacronym{SIMO}{SIMO}{single-input multiple-output}
\newacronym{MISO}{MISO}{multiple-input single-output}
\newacronym{AoD}{AoD}{angle-of-departure}
\newacronym{AoA}{AoA}{angle-of-arrival}
\newacronym{RF}{RF}{radio frequency}
\newacronym{SIM}{SIM}{stacked intelligent metasurfaces}
\newacronym{FPGA}{FPGA}{field programmable gate array}
\newacronym{UPA}{UPA}{uniform planar array}
\newacronym{CC}{CC}{communication-centric}
\newacronym{I/O}{I/O}{input-output}
\newacronym{iid}{i.i.d.}{independent and identically distributed}
\newacronym{IoT}{IoT}{internet of things}
\newacronym{V2X}{V2X}{vehicle-to-everything}
\newacronym{NTN}{NTN}{non-terrestrial network}
\newacronym{LEO}{LEO}{low-earth orbit}
\newacronym{THz}{THz}{terahertz}
\newacronym{EM}{EM}{electromagnetic}
\newacronym{RIS}{RIS}{reconfigurable intelligent surface}
\newacronym{DoA}{DoA}{direction-of-arrival}
\newacronym{DD}{DD}{doubly-dispersive}
\newacronym{ODDM}{ODDM}{orthogonal delay-Doppler division multiplexing}
\newacronym{LoS}{LoS}{line-of-sight}
\newacronym{NLoS}{NLoS}{non-line-of-sight}
\newacronym{6G}{6G}{sixth generation}
\newacronym{MPDD}{MPDD}{metasurfaces-parametrized DD}
\newacronym{GaBP}{GaBP}{Gaussian Belief Propagation}
\newacronym{MSE}{MSE}{mean-squared-error}
\newacronym{sIC}{soft IC}{soft interference cancellation}
\newacronym{soft RG}{soft RG}{soft replica generation}
\newacronym{BG}{BG}{belief generation}
\newacronym{SGA}{SGA}{scalar Gaussian approximation}
\newacronym{CLT}{CLT}{central limit theorem}
\newacronym{PDF}{PDF}{probability density function}
\newacronym{QPSK}{QPSK}{quadrature phase-shift keying}
\newacronym{OQAM}{OQAM}{offset quadrature amplitude modulation}
\newacronym{LMMSE}{LMMSE}{linear minimum mean square error}
\newacronym{SNR}{SNR}{signal-to-noise ratio}
\newacronym{OOBE}{OOBE}{out-of-band emissions}
\newacronym{PAPR}{PAPR}{peak-to-average power ratio}
\newacronym{AFBM}{AFBM}{Affine Filter Bank Modulation}
\newacronym{FBMC}{FBMC}{Filter Bank Multicarrier Modulation}
\newacronym{PPN}{PPN}{polyphase network}
\newacronym{SIR}{SIR}{signal-to-interference ratio}

\newcommand\scalemath[2]{\scalebox{#1}{\mbox{\ensuremath{\displaystyle #2}}}}

\begin{document}

%

%
%
\title{Affine Filter Bank Modulation:\\ \!\!A New Waveform for High Mobility Communications\!\!}

\author{\IEEEauthorblockN{Henrique L. Senger\textsuperscript{\orcidlink{0009-0004-1586-8168}}\IEEEauthorrefmark{1}, Gustavo P. Gonçalves\textsuperscript{\orcidlink{0009-0000-8260-4390}}\IEEEauthorrefmark{1},  Bruno S. Chang\textsuperscript{\orcidlink{0000-0003-0232-7640}}\IEEEauthorrefmark{1}, \\ 
Hyeon Seok Rou\textsuperscript{\orcidlink{0000-0003-3483-7629}}\IEEEauthorrefmark{2},  Kuranage Roche Rayan Ranasinghe\textsuperscript{\orcidlink{0000-0002-6834-8877}}\IEEEauthorrefmark{2}, \\
Giuseppe Thadeu Freitas de Abreu\textsuperscript{\orcidlink{0000-0002-5018-8174}}\IEEEauthorrefmark{2} and Didier Le Ruyet\textsuperscript{\orcidlink{0000-0002-9673-2075}}\IEEEauthorrefmark{3}} \vspace{1ex}

\IEEEauthorblockA{
\IEEEauthorrefmark{1}CPGEI/Electronics Department, Federal University of Technology - Paraná, Curitiba, Brazil \\[0.15ex]
\IEEEauthorrefmark{2}School of Computer Science and Engineering, Constructor University, Bremen, Germany \\[0.15ex]
\IEEEauthorrefmark{3}CEDRIC/Conservatoire National des Arts et Métiers - Paris, France \\
}
\vspace{-4ex}
}

\maketitle

\begin{abstract} 
%
%
We propose a new waveform suitable for \ac{ISAC} systems facing \ac{DD} channel conditions, as typically encountered in high mobility scenarios.
Dubbed \ac{AFBM}, this novel waveform is designed based on a filter-bank structure, known for its ability to suppress \ac{OOBE}, while integrating a \ac{DAFT} precoding stage which yields low \ac{PAPR} and robustness to \ac{DD} distortion, as well as other features desirable for \ac{ISAC}.
Analytical and simulation results demonstrate that \ac{AFBM} maintains quasi-orthogonality similar to that of \ac{AFDM} in \ac{DD} channels, while achieving \ac{PAPR} levels 3 dB lower, in addition to \ac{OOBE} as low as -100 dB when implemented with PHYDYAS prototype filters.
\end{abstract}

\IEEEpeerreviewmaketitle

\glsresetall

\section{Introduction}

Next-generation wireless systems are expected to not only enable higher data rates, increase reliability, and provide low latency, but also support many emerging applications, from high-mobility communications \cite{Jingxian2016} to \ac{ISAC} \cite{Zhiqing_IoT2023, NuriaISAC2024, RanasingheTWC2025}.
High mobility scenarios and/or the higher carrier frequencies required to that end, however, imply the presence of \ac{DD} channels, which are known \cite{Svante2007} to be challenging to \ac{OFDM}, the standard waveform used in most modern communications systems, since the large Doppler shifts present in these channels lead to severe inter-carrier interference and consequent link performance degradation.

To overcome this challenge, several novel waveforms have been proposed in the last years \cite{Rou_SPM_2024}, with \ac{AFDM} being one of the most recent and promising solutions.
The \ac{AFDM} \cite{bemani2023} waveform maps data symbols into chirp-based subcarriers in the \ac{DAF} domain, and is capable of attaining full diversity in \ac{DD} channels via the adjustment of chirp parameters, which are tuned based on the channel characteristics.
However, \ac{AFDM} suffers from a high \ac{PAPR} and poor spectral containment, leading to challenges in implementation such as inefficient power amplification and high \ac{OOBE}.

Significant effort has been made recently, therefore, to address these limitations.
In \cite{tao2024daft}, for instance, an access scheme was proposed in which the \ac{DFT} operation of  a spread-\ac{OFDM} system (DFT-s-OFDMA) was replaced by a \ac{DAFT}, yielding a spread-\ac{AFDM} access (DAFT-s-AFDMA) approach shown to have improved \ac{PAPR} properties.
In turn, in \cite{omar2020spectrum} a modified version of \ac{OCDM} dubbed c-\ac{OCDM} was presented, where spectral containment is improved by applying zero-padding in the frequency domain on the chirp matrices prior to modulation.
And given that \ac{AFDM} and \ac{OCDM} differ mainly on the choice of chirp parameters, the spectral containment approach of \cite{omar2020spectrum}, it is natural that a similar idea was presented for AFDM in \cite{savaux2024}, where a modified transceiver structure to reduce the overall computational complexity is also shown.

Approaching the problem from the perspective of \ac{OOBE}, on the other hand, \ac{FBMC} emerges as an alternative to design \ac{DD}-robust \ac{ISAC} waveforms, due to the per-subcarrier filtering process inherent of filter-bank methods. 
And while regular \ac{FBMC} systems are limited to real orthogonality and \ac{OQAM} modulation to fulfill the Balian-Low theorem, precoded versions such as those proposed in \cite{zakaria2012novel, nissel2018pruned, pereira2021novel, pereira2022generalized} allow the restoration of complex orthogonality, with additional low \ac{PAPR} features obtained via \ac{DFT}-pruning.

In spite of all the aforementioned efforts, the fact is that no \ac{ISAC}-enabling waveform has been proposed so far that can simultaneously deliver robustness to \ac{DD} and low \ac{PAPR} and \ac{OOBE}.
We therefore introduce in this work a novel waveform -- referred to as the \ac{AFBM} -- which is designed to address the limitations of existing alternatives all at once.
To that end, our proposed waveform in essence integrates a pruned discrete \ac{DAFT} precoding stage and \ac{FBMC} filtering to the \ac{AFDM} scheme.
In particular, \ac{AFBM} achieves low \acp{PAPR} close to that of DAFT-s-AFDM, good \ac{OOBE} characteristics similar to that of \ac{FBMC}, and quasi-orthogonality in \ac{DD} channels comparable to that of \ac{AFDM}.

These claims are demonstrated both via analytical and simulation results, with the remainder of the paper organized as follows.
First, the proposed model and the design of \ac{AFBM} are described in Section \ref{secSysModel}.
In Section \ref{secSymResults} simulation results are presented, comparing the performance of the scheme against \ac{AFDM} in terms of \ac{BER}, \ac{PAPR}, and \ac{OOBE}.
Finally, future works and conclusion are offered in Section \ref{secConclusion}.


\section{System Model and Proposed AFBM Design}
\label{secSysModel}

\subsection{Transceiver Structure}

In a filter bank-based multicarrier system, filtering is performed on each subcarrier, providing a well-localized spectrum, significant rejection of \ac{OOBE}, and the suppression of interference between symbols and between subcarriers. 

At the processing stage through the transmitter filter bank, symbols are mapped onto a grid with $L$ subcarriers and $K$ time indices. 
Then, in order to maintain complex orthogonality, and since half of the system's $L$ subcarriers must be reserved as a guard interval to avoid filter interference, transmission is carried out at a double rate with respect to the conventional transmission seen in \ac{OFDM}/\ac{AFDM} systems, namely, every $L/2$ samples.


To facilitate analysis and for clarity of exposition, we will represent the considered system using matrix notation.
Let us then define the $n$-point \ac{DFT} matrix $\mathbf{F}_n \in \mathbb{C}^{n \times n}$ as
\begin{eqnarray}
\mathbf{F}_n = \big\{e^{j2\pi kl/n} \big\}^{n-1}_{k,l = 0}.
\label{DFTmatrix}
\end{eqnarray}

The considered filter bank-based system can be characterized as a block structure composed of $K$ symbols with a duration of $T/2$ seconds, where each symbol has $L$ subcarriers spaced by $F$ Hz.
Thus, in the TF domain we obtain a grid with $L$ points spaced by $F$ Hz along the frequency axis and $K$ points along the time axis with a space of $T/2$ seconds.
In this sense, the total bandwidth is defined as $B=LF$ and the total transmission time interval as $KT/2$. 

Next, let $\mathbf{A} \in \mathbb{C}^{L \times K}$ be a matrix containing the symbols in the time-frequency domain.
The data symbols are inserted in the first and last $L/4$ positions of the grid according to a pre-established transmission strategy in order to avoid interference from the filter bank (this will be detailed later in the Section), such that the signal after precoding can be expressed as
\begin{eqnarray}
\mathbf{X} = \mathbf{W}_{L} \textrm{diag}\{\tilde{\mathbf{b}}\} \mathbf{A},
\label{OTFSFB}
\end{eqnarray}
where $\mathbf{W}_{L} \in \mathbb{C}^{L \times L}$ represents the $L$-point DAFT matrix, given by
\begin{equation}
\mathbf{W}_{L} = \mathbf{\Lambda}_{c_1,L}\mathbf{F}_{L}\mathbf{\Lambda}_{c_2,L},
\end{equation}
where $\mathbf{\Lambda}_{c_i,L} \in \mathbb{C}^{L \times L} = \text{diag}[e^{-j2\pi c_i (0)^2} , \dots, e^{-j2\pi c_i (L-1)^2}]$ is a $L$-size diagonal chirp matrix with a central digital frequency of $c_i$ while $\tilde{\mathbf{b}} \in \mathbb{C}^{L \times 1}$ is the filter compensation vector. 

We emphasize that $\tilde{\mathbf{b}}$ is needed for transceiver complex orthogonality and will also be detailed later in this Section.
%
In addition, the length $P$ of the \ac{IDAFT} at the transmitter must be greater than $L$ to prevent the precoding stage \ac{DAFT} from being nullified by the \ac{IDAFT} of the filter bank structure.

Finally, the length $P$ must be smaller than the size of the filter bank $N$ so that the chirps are sampled at a rate lower than Nyquist's, allowing frequency containment.
From all the above, it follows that the \ac{DAFT} $\mathbf{\tilde{W}}_{P} \in  \mathbb{C}^{L \times P}$ can be expressed as
\begin{eqnarray}
\mathbf{\tilde{W}}_{P} = \begin{bmatrix}
\mathbf{I}_L &  \mathbf{0}_{L\times (P-L)}  \end{bmatrix}
\mathbf{W}_P.
\label{dft_espalhada}
\end{eqnarray}


Afterwards, by using frequency domain zero padding, the output matrix $\mathbf{Q}_{P}$ for a given block can be generated as
\begin{equation}
\mathbf{Q}_{P} = \mathbf{F}_{N}^{H}\mathbf{T}\mathbf{F}_P \mathbf{\tilde{W}}^H_{P},
\label{eq:Q_P}
\vspace{-0.5ex}
\end{equation}
where $\mathbf{T}\triangleq \begin{bmatrix} \mathbf{I}_{P,l}^{T} & \mathbf{0}_{(N-P)\times P} & \mathbf{I}_{P,u}^{T} \end{bmatrix}$,
with $\mathbf{T}^{T}\mathbf{T} = \mathbf{I}_P$, is an $N \times P$ matrix, with $\mathbf{I}_{P,u}$ and $\mathbf{I}_{P,l}$ denoting the first and last $P/2$ columns of the $P \times P$ identity matrix $\mathbf{I}_P$, respectively.

Considering the transmission of $K$ blocks, the block matrix $\mathbf{Q} \in  \mathbb{C}^{NK\times LK}$ is defined as
\begin{eqnarray}
\mathbf{Q} =  \mathbf{I}_{K} \otimes \mathbf{Q}_{P} , 
\vspace{-0.5ex}
\end{eqnarray}
where $\otimes$ refers to the Kronecker product which maps $\mathbf{Q}_{P}$ to the correct time positions.

Then, the transmitted data can be obtained by convolving the output with the prototype filter impulse response through a Toeplitz filter matrix.
Let us consider the diagonal matrix $\mathbf{G}_p$ corresponding to the filter coefficients,  that is, $\mathbf{G}_p =$ diag$(\mathbf{g}_p) \in  \mathbb{R}^{N/2 \times N/2}$ for $p = 0,1,2,...,2 O-1$, where $\mathbf{g}_{p}$ is given by  $\mathbf{g}_p = [g[pN/2],g[pN/2+1],   \ldots , g[pN/2+N/2-1]]$ and $\mathbf{g}$ is the prototype filter of length $ON$, where $O$ is the filter overlap factor.
Thus, the block Toeplitz filter matrix $\mathbf{G} \in  \mathbb{R}^{ON + (K-1)N/2 \times NK}$ can be generated from the vector $[ \mathbf{\tilde{G}}_0  \hspace{0.1cm}  \mathbf{\tilde{G}}_1   \hspace{0.1cm} \mathbf{\tilde{G}}_2 \ldots \mathbf{\tilde{G}}_{2O -1}]^T$, namely
\begin{eqnarray}
\mathbf{G} = 
\scalemath{0.8}{\begin{bmatrix}
\mathbf{G}_0  & \mathbf{0} & \mathbf{0} & \mathbf{0} & \ldots  & \mathbf{0} \\
\mathbf{0} & \mathbf{G}_1  & \mathbf{G}_0  & \mathbf{0} & \ldots  & \mathbf{0} \\
\mathbf{G}_2  & \mathbf{0} & \mathbf{0} & \mathbf{G}_1  &  \ldots  & \mathbf{0} \\
 \mathbf{0} & \mathbf{G}_3  & \mathbf{G}_2 & \mathbf{0} & \ldots  & \mathbf{0} \\
 \vdots &  \mathbf{0} &  \mathbf{0} & \mathbf{G}_3  &  \ldots  & \mathbf{0} \\
 \vdots &  \vdots &  \vdots &  \vdots &  \ddots  &   \vdots \\
 \mathbf{G}_{2O-4} & \vdots &  \vdots &  \vdots & \ddots  &  \mathbf{0} \\
 \mathbf{0} & \mathbf{G}_{2O-3} & \mathbf{G}_{2O-4} &  \vdots & \ddots & \mathbf{G}_1 \\
 \mathbf{G}_{2O-2} &  \mathbf{0} &  \mathbf{0} &  \mathbf{G}_{2O-3} & \ddots &  \mathbf{0} \\
 \mathbf{0} & \mathbf{G}_{2O-1} & \mathbf{G}_{2O-2} & \mathbf{0} & \ddots &  \mathbf{G}_3 \\
 \vdots &  \mathbf{0} & \mathbf{0} & \mathbf{G}_{2O-1} & \ddots & \vdots \\
  \vdots &  \vdots & \ddots &   \vdots &  \ddots &  \mathbf{0} \\
   \mathbf{0} &   \mathbf{0} &  \ldots & \mathbf{0} &  \ddots & \mathbf{G}_{2O-1} 
\end{bmatrix}}.
\label{fhddg}
\end{eqnarray}

Once the transmission matrix  has been defined, the output data vector $\mathbf{s}$ of length $M = ON + \frac{N}{2}(K-1)$ is given by
\begin{eqnarray}
\mathbf{s} =  	\mathbf{G}\mathbf{Q}\mathbf{x} = \mathbf{\bar{G}}\mathbf{x},
\label{transmissao2}
\end{eqnarray}
where $\mathbf{\bar{G}} = \mathbf{G} \mathbf{Q} \in \mathbb{C}^{M \times LK}$ and $\mathbf{x}  = \textrm{vec}(\mathbf{X}) \in \mathbb{C}^{LK \times 1}$.

The symbols are delayed from each other every $N/2$ samples, which the $\mathbf{G}$ structure itself does transparently with the inclusion of matrices $\mathbf{0}_{N/2}$. The representation as a sum of matrices delayed among themselves can be seen in \cite{pereira2022generalized}.

\begin{figure*}[h]
\centering
\includegraphics[width=1\textwidth]{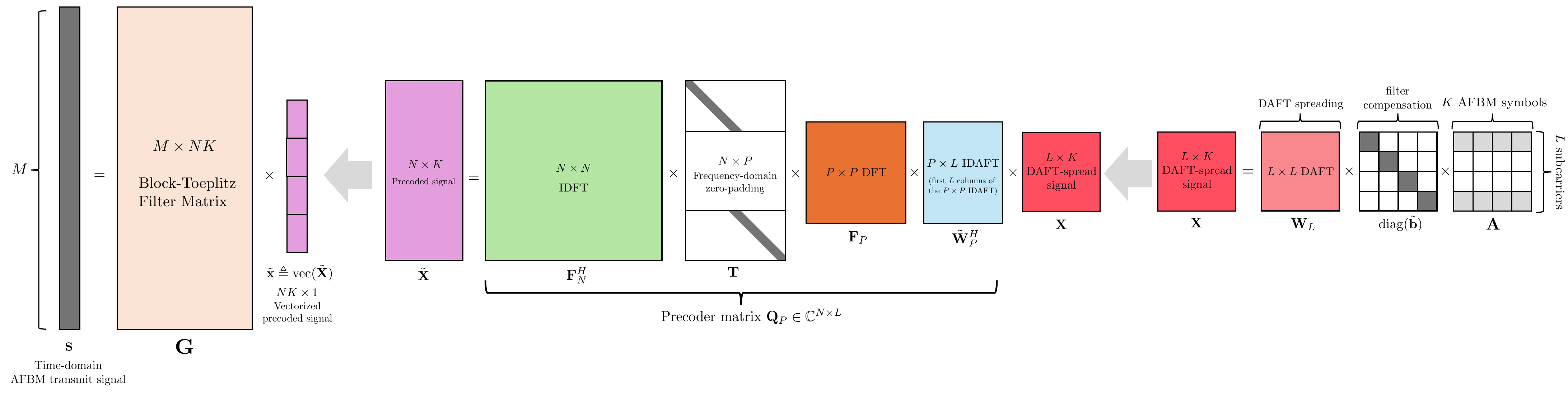}
\vspace{-4ex}
\caption{Illustration of the full AFBM modulation framework.}
\label{fig:AFBMmod}
\end{figure*}

It can be seen from the above that the proposed structure is an affine-precoded filter bank scheme, which can be implemented efficiently using a \ac{PPN} together with an \ac{IFFT}.
Another interpretation is that in the proposed waveform the standard AFDM sinc chirp subcarriers were replaced with chirp filtered ones.
The full modulation schematic diagram of the proposed AFBM waveform has been illustrated in Figure \ref{fig:AFBMmod}.

The signal vector $\mathbf{s}$ is transmitted over a time-varying multipath channel, i.e., the doubly-dispersive channel, described concisely by the circular convolutional matrix form $\mathbf{H} \in  \mathbb{C}^{M \times M}$ \cite{nissel2018pruned}. Consequently, the received signal $\mathbf{r} \in  \mathbb{C}^{M \times 1}$  is expressed by $\mathbf{r} = \mathbf{H}\mathbf{s} + \mathbf{n}$,
where  $\mathbf{n}$ represents additive white Gaussian noise (AWGN) samples with zero mean and power $\sigma_n^2$. At the receiver, the opposite process to that at the transmitter is carried out. 

%

Firstly, the received signal is demodulated by $\mathbf{\bar{G}}^H$, generating $\mathbf{\tilde{x}}$ $\in \mathbb{C}^{LK \times 1}$.
Then (and considering all time instants), the detected symbols represented by $\mathbf{\tilde{A}} \in C^{L \times K}$ are obtained through the \ac{IDAFT} combined with the compensation stage
\begin{eqnarray}
\mathbf{\tilde{A}} = \mathbf{W}^H_L \textrm{diag}\{\tilde{\mathbf{b}}\} \mathbf{\tilde{X}}
\label{12ghtj}
\end{eqnarray}
where $\mathbf{\tilde{X}} = [\mathbf{\tilde{x}}_0;\mathbf{\tilde{x}}_1;\hdots;\mathbf{\tilde{x}}_{K-1}] \in C^{L \times K}$.

In order to obtain the estimated symbols, the $L/2$ intermediate values between the first and last $L/4$ symbols are discarded, since no data was transmitted in those positions in $\mathbf{A}$. For symmetry reasons, the compensation stage is used in both the receiver and transmitter. However, it can only be applied to one side effectively.


\subsection{Restoration of Complex Orthogonality}

It is possible obtain a filter bank waveform using well localized filters and with complex orthogonality by using the \ac{DAFT} together with the filter compensation stage to cancel interference. Thus, this part of the precoding  presented previously in~\eqref{OTFSFB} is defined as
\begin{eqnarray}
\mathbf{C}_f = \mathbf{W}_{L} \textrm{diag}\{\tilde{\mathbf{b}}\},
\label{ferf44}
\end{eqnarray}
where $\mathbf{C}_f \in \mathbb{C}^{L \times L}$ represents the process within the precoding responsible for restoring complex orthogonality and consists of an $L$-point \ac{DAFT} combined with the compensation multiplicative factor.

It follows that in order to preserve complex orthogonality, $\mathbf{C}_f$ must be selected to guarantee the following condition, $i.e.$
\begin{eqnarray}
\mathbf{C}^H_f\mathbf{Q}_{P}^{H}\mathbf{\widetilde{G}}^T\mathbf{\widetilde{G}}\mathbf{Q}_{P} \mathbf{C}_f \approx 
\mathbf{U}, 
\label{eq:complex_orthogonality}
\end{eqnarray}
where $\mathbf{U} \in \mathbb{R}^{L \times L}$ is a matrix with unit values in the first and last $L/4$ positions of its main diagonal and zeros in the rest. $\mathbf{\widetilde{G}}$ represents the filtering corresponding to the transmission of a single multicarrier symbol, expressed by $\mathbf{\widetilde{G}} = [\mathbf{G}_0; \mathbf{G}_1;...;\mathbf{G}_{2 O-1}] \in \mathbb{R}^{ON \times N}$. 

By substituting~\eqref{ferf44} into~\eqref{eq:complex_orthogonality}, the $\tilde{l}$-th position of $\tilde{\mathbf{b}}$ can be expressed by
\begin{eqnarray}
[\mathbf{\tilde{b}}]_{\tilde{l}} = \left \{ \scalemath{0.8}{\begin{matrix} \sqrt{\frac{1}{[\mathbf{\tilde{c}}]_{\tilde{l}}}},&\mbox{for } \tilde{l} = \left[ 0,\ldots,\frac{L}{4}-1 ; L-\frac{L}{4}-1,\ldots,L-1 \right] \\ 0, & \mbox{otherwise} \end{matrix}} \right. 
\end{eqnarray}
with
\begin{eqnarray}
\mathbf{\tilde{c}} &=& \textrm{diag}\{\mathbf{W}^H_L\mathbf{Q}_{P}^{H}\mathbf{\widetilde{G}}^T\mathbf{\widetilde{G}}\mathbf{Q}_{P}\mathbf{W}_L\}.
\end{eqnarray}

The compensation stage comprises a multiplicative factor that compensates for the interference in the transmitted symbol that stems from the filter coefficient. The filter coefficients are known, since they come from a pre-established prototype filter \cite{pereira2023two}. Thus, the objective is to ensure that the filter interference is limited to only one coefficient so that the compensation is done correctly. This is guaranteed by using an overlap factor ($O$) less than or equal to $1.5$ \cite{pereira2022generalized}. If the overlap factor is higher than $1.5$ the overall response in~\eqref{eq:complex_orthogonality} will have interference components outside the main diagonal, which will reduce the \ac{SIR}.

\subsection{AFBM Effective Channel Analysis}

In this section, the effective channel of the AFBM waveform is derived to analyze its robustness in the doubly-dispersive channel, as was proposed for the recently discovered AFDM and OTFS waveforms \cite{bemani2023,Rou_SPM_2024}.

To that end, first consider the transmit symbols $\mathbf{A} \in \mathbb{C}^{L \times K}$ reformulated into its vectorized form $\mathbf{a} \in \mathrm{vec}(\mathbb{A})\in \mathbb{C}^{L K \times 1}$.
Then, vectorized form of the DAFT-spread transmit signal $\mathbf{x} \in \mathbb{C}^{LK \times 1}$ is given by
\begin{align}
    \mathbf{x} & \triangleq \mathrm{vec}(\mathbf{X}) = \mathrm{vec}\big(\overbrace{\mathbf{W}_L\mathrm{diag} (\tilde{\mathbf{b}})}^{\triangleq \mathbf{C}_f \, \in \, \mathbb{C}^{L \times L}}\mathbf{A}\big) \in \mathbb{C}^{LK \times 1} \\
    &  = \underbrace{\big(\mathbf{I}_K \otimes \mathbf{C}_f \big)}_{\triangleq \mathbf{C} \, \in \, \mathbb{C}^{LK \times LK}}\!\!\cdot \,\mathrm{vec}(\mathbf{A}) =  \mathbf{C} \cdot \mathbf{a}. \nonumber
\end{align}

Then, following eq. \eqref{transmissao2}, the full AFBM transmit signal in time-domain can be efficiently described in terms of the filter matrix $\mathbf{G}$, modified \ac{IDAFT} matrix $\mathbf{Q}_P$, and the compensation matrix $\mathbf{C}_f$ leveraging identities involving Kronecker products
\begin{align}
    \mathbf{s} & = \mathbf{G} \mathbf{Q} \mathbf{C} \mathbf{a}  = \mathbf{G} \big(\mathbf{I}_{K} \otimes \mathbf{Q}_{P}\big) \cdot \big(\mathbf{I}_K \otimes \mathbf{C}_f \big) \mathbf{a} \in \mathbb{C}^{M \times 1} \nonumber \\
    & = \mathbf{G} \big(\mathbf{I}_{K} \otimes \mathbf{Q}_{P} \mathbf{C}_f \big) \mathbf{a}.
\end{align}

Then, the received signal is given by
\begin{equation}
    \mathbf{r} \triangleq \underbrace{\Big(\sum_{r=1}^{R}h_r \mathbf{\Phi}_r\mathbf{Z}^{f_r} \mathbf{\Pi}^{\ell_r}\Big)}_{\triangleq \mathbf{H}} \mathbf{G} \Big(\mathbf{I}_{K} \otimes \mathbf{Q}_{P} \mathbf{C}_f)\Big) \mathbf{a},
\end{equation}
where $\mathbf{H} \in \mathbb{C}^{M \times M}$ is the doubly-dispersive wireless channel described by $R$ resolvable propagation paths, where each $r$-th scattering path induces a delay $\tau_r \in [0, \tau^\mathrm{max}]$ and Doppler shift $\nu_r \in [-\nu^\mathrm{max}, +\nu^\mathrm{max}]$ to the received signal, where $\ell_r \triangleq \lfloor \frac{\tau_r}{T_\mathrm{s}} \rceil \in \mathbb{N}_0$ and $f_r \triangleq \frac{N\nu_r}{f_\mathrm{s}} \in \mathbb{R}$ are the normalized integer path delay and normalized digital Doppler shift of the $r$-th propagation path, with sampling frequency $f_\mathrm{s} \triangleq \frac{1}{T_\mathrm{s}}$.

As elaborated in \cite{Rou_SPM_2024}, each $r$-th path of the full circular convolutional matrix $\mathbf{H} \triangleq \sum_{r=1}^{R}h_r \mathbf{\Phi}_{r} \mathbf{Z}^{f_r} \mathbf{\Pi}^{\ell_r} \in \mathbb{C}^{M \times M}$, is parametrized by the complex channel fading coefficient $h_r \in \mathbb{C}$, the diagonal prefix phase matrix $\boldsymbol{\Phi}_r  \!\in\! \mathbb{C}^{M \times M}$ with the IDAFT-based chirp-cyclic prefix phase function $\phi(m) \triangleq  c_1(M^2 - 2Mm) $, given by
\begin{equation}
\color{black}
\!\!\!\mathbf{\Phi}_{r} \!\triangleq\! \mathrm{diag}\big[e^{-j2\pi \cdot \phi(\ell_r)}, \ldots, e^{-j2\pi \cdot\phi(1)}, 1, \ldots, 1\big] \!\in\! \mathbb{C}^{M \times M}\!,\!\!\!
\label{eq:CCP_phase_matrix}
\end{equation}
the diagonal roots-of-unity matrix $\mathbf{Z} \!\in\! \mathbb{C}^{M \times M}$ given by 
\begin{equation}
\mathbf{Z} \triangleq \mathrm{diag}\big[e^{-j2\pi\frac{(0)}{M}}, \,\ldots\,, e^{-j2\pi\frac{(M\!-\!1)}{M}}\big] \in \mathbb{C}^{M \times M},
\label{eq:Z_matrix}
\end{equation}
which is taken to the $f_r$-th power, and the right-multiplying circular left-shift matrix $\mathbf{\Pi} \in \mathbb{C}^{M \times M}$.
In all, this channel matrix composes of $R$ shifted diagonals whose position is dependent on the normalized path delay, and coefficients dependent on the normalized Doppler shift of the path.

\begin{figure}[t!]
	\centering		
	\includegraphics[width=0.8\columnwidth]{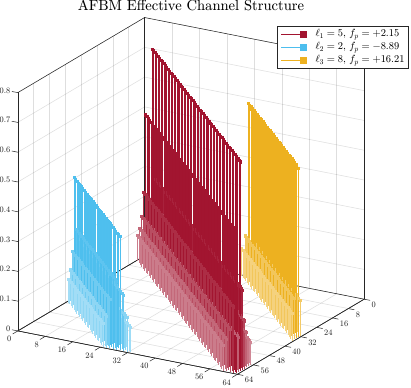}
		\caption{3D illustration of a three-path AFBM effective channel of the DAFT-spread signal, where the normalized delay and Doppler shifts are denoted in legend.
        The AFBM parameters are $L = 64, N = 128, P = 128, O = 4$ with the PHYDYAS4 prototype filter.}
	\label{fig:AFBM_chan}
\end{figure}

For the sake of analyzing the properties of the effective channel with respect to the double dispersitivity, i.e., the effect of delay and Doppler of each path unto the channel structure, let us simplify without loss of generality, by considering only one AFBM symbol, $K = 1$, such that the Kronecker products with the identity matrix $\mathbf{I}_K$ are reduced.

Furthermore, the effective channel $\mathbf{H}_\mathrm{eff}$ will be defined as the full transformation matrix only between the \ac{DAFT}-spread signal $\mathbf{x}$ and the demodulated signal $\tilde{\mathbf{x}}$, i.e., without de-spreading and filter compensation, as these operations do not alter the properties of the waveform over the core doubly-dispersive channel operations by the modulating DAFT/IDAFT $\mathbf{Q}_P$.

Then, including the corresponding demodulation and equalization, 
the effective doubly-dispersive channel matrix of the AFBM waveform is defined as 
\begin{equation}
    \mathbf{H}_\mathrm{eff} \triangleq \mathbf{Q}_P^{{H}} \mathbf{G}^{H} \Big(\sum_{r=1}^{R}h_r \mathbf{\Phi}_r\mathbf{Z}^{f_r} \mathbf{\Pi}^{\ell_r}\Big)  \mathbf{G} \mathbf{Q}_{P} \in \mathbb{C}^{L \times L},
\end{equation}
which captures the structure of the doubly-dispersive channel after the filter shaping and the truncated IDAFT/DAFT.

This AFBM effective channel has been illustrated in Fig. \ref{fig:AFBM_chan} for a scenario with three distinct propagation paths, where the spreading of the band-diagonals of the effective channel is observed.
This behavior is alike to the \ac{AFDM} waveform \cite{bemani2023,Rou_SPM_2024} inherently due to the IDAFT-based modulation process -- except for the different sideband behavior from the filter design -- resulting in a deterministic shift of the diagonals based on the path delay and Doppler shift indices, which is the key factor enabling the robustness of the waveform by mitigating inter-path interference, in addition to the highly beneficial implications to \ac{ISAC} parameter estimation.
In other words, the proposed AFBM enables robust communications over doubly-dispersive channels alike to the AFDM.

\section{Simulation Results}
\label{secSymResults}

In this section, simulation results presenting \ac{PAPR} and \ac{OOBE} of the proposed scheme will be shown, together with a comparison taking into account the classical \ac{AFDM} scheme.
For these simulations, the total number of subcarriers $L$ is 128, the chirp size $P$ is 192 and the filter bank \ac{DFT} size $N$ is 256.

Each transmission was composed of $K = 8$ symbols, with a carrier frequency $f_c$ of 4 GHz.
The channel is a doubly dispersive one, with three resolvable paths and corresponding normalized delays and digital Doppler shifts.
We recall that the chirp frequencies for each (I)DAFT were chosen to uphold the orthogonality condition \cite{Rou_SPM_2024} $2(f^{\text{max}} + \xi)(\ell^{\text{max}}+1) + \ell^{\text{max}} \leq P$, where  $f^{\text{max}}$ and $\ell^{\text{max}}$ are, respectively, the maximum normalized digital Doppler shift and delay of the channel and $\xi \in \mathbb{N}_0$ is a free parameter determining the so-called guard width, denoting the number of additional guard elements around the diagonals to anticipate for Doppler-domain interference.

\begin{figure}[t]
  \centering
  \includegraphics[width=\columnwidth]{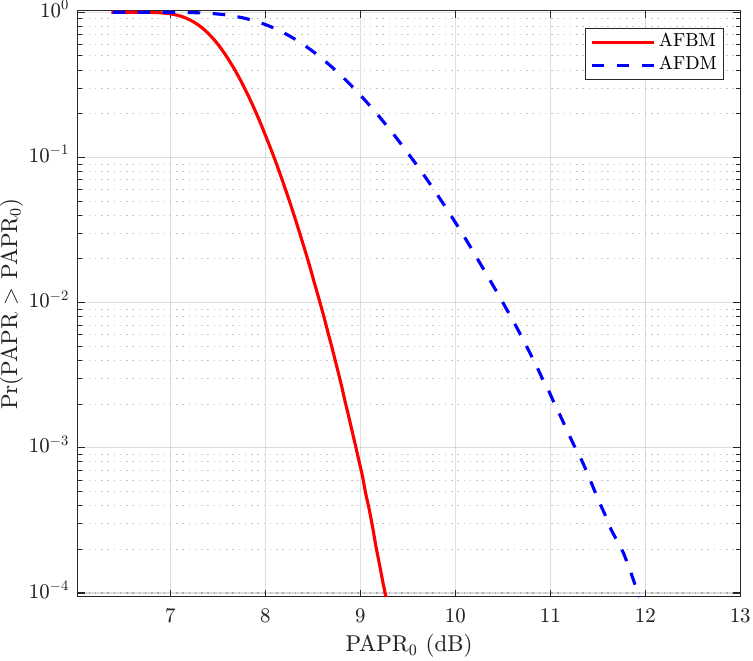}%
  \caption{\ac{PAPR} performance of \ac{AFDM} and \ac{AFBM}.}
  \label{fig:afdm_papr}
\end{figure}

Figure~\ref{fig:afdm_papr} presents \ac{PAPR} results from the considered systems. It can be seen that, due to its single-carrier like structure, the proposed structure has an advantage of 2 dB with respect to regular \ac{AFDM}, which has the same high \ac{PAPR} of regular \ac{OFDM}-based schemes. Also, since the proposed scheme can be seen as an affine version of the pruned \ac{DFT} spread \ac{FBMC}, the low \ac{PAPR} characteristic of the original version is maintained.

\begin{figure}[t]
  \centering
  \includegraphics[width=\columnwidth]{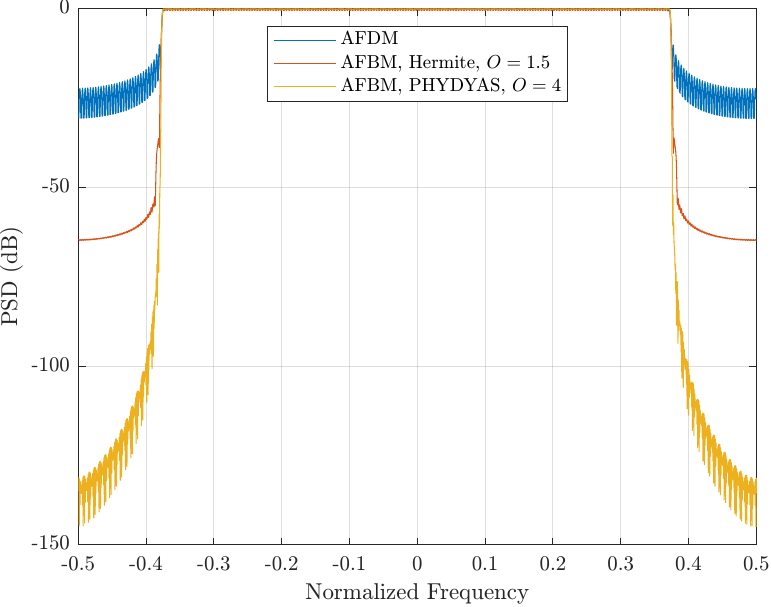}%
  \caption{\ac{OOBE} Performance of \ac{AFDM} and \ac{AFBM} with the Hermite and PHYDYAS prototype filters.}
  \label{fig:afdm_oob}
\vspace{-2ex}
\end{figure}

The \ac{OOBE} of the considered schemes are shown in Figure~\ref{fig:afdm_oob}. In this scenario, two prototype filters were considered for \ac{AFBM}: the truncated Hermite  (with $O = 1.5$) and the full PHYDYAS (with $O = 4$). Due to the usage of well-localized filters instead of the rectangular window there is a significant advantage in \ac{OOBE} of the \ac{AFBM} system with respect to regular \ac{AFDM}. Here, an interesting compromise regarding the prototype filter choice appears: if lower overall latency is prioritized a shorter filter can be selected, at the expense of worse (but still much better than \ac{AFDM}) spectral localization. Finally, we remember that if the nonlinear characteristics of high power amplifiers are considered, the good \ac{OOBE} characteristics of the proposed waveform will not be severely affected due to the low \ac{PAPR}.

\section{Conclusion}
\label{secConclusion}

We introduced an affine-precoded filter bank scheme called \ac{AFBM} in this work. It was seen that it has lower \ac{PAPR} and \ac{OOBE} than regular \ac{AFDM}. In the follow-up, new detection structures taking into account the interference introduced by prototype filters with high overlap factors and simplified structures to reduce the overall computational complexity will be proposed. Moreover, the sensing performance of this scheme will be investigated. Furthermore, adapted channel estimation schemes for the proposed waveform will be proposed, as well as studies dealing with its robustness to high mobility conditions. Finally, chirp parameters optimization to reduce even further the \ac{PAPR} will be developed.

\balance

\selectlanguage{english}

\begin{thebibliography}{10}
\providecommand{\url}[1]{#1}
\csname url@samestyle\endcsname
\providecommand{\newblock}{\relax}
\providecommand{\bibinfo}[2]{#2}
\providecommand{\BIBentrySTDinterwordspacing}{\spaceskip=0pt\relax}
\providecommand{\BIBentryALTinterwordstretchfactor}{4}
\providecommand{\BIBentryALTinterwordspacing}{\spaceskip=\fontdimen2\font plus
\BIBentryALTinterwordstretchfactor\fontdimen3\font minus
  \fontdimen4\font\relax}
\providecommand{\BIBforeignlanguage}[2]{{%
\expandafter\ifx\csname l@#1\endcsname\relax
\typeout{** WARNING: IEEEtran.bst: No hyphenation pattern has been}%
\typeout{** loaded for the language `#1'. Using the pattern for}%
\typeout{** the default language instead.}%
\else
\language=\csname l@#1\endcsname
\fi
#2}}
\providecommand{\BIBdecl}{\relax}
\BIBdecl

\bibitem{Jingxian2016}
J.~Wu and P.~Fan, ``A survey on high mobility wireless communications:
  Challenges, opportunities and solutions,'' \emph{IEEE Access}, vol.~4, pp.
  450--476, 2016.

\bibitem{Zhiqing_IoT2023}
Z.~Wei \emph{et. al},
  ``Integrated sensing and communication signals toward {5G-A and 6G}: A
  survey,'' \emph{IEEE Internet of Things Journal}, vol.~10, no.~13, 2023.

\bibitem{NuriaISAC2024}
N.~González-Prelcic  \emph{et. al}, ``The
  integrated sensing and communication revolution for {6G}: Vision, techniques,
  and applications,'' \emph{Proceedings of the IEEE}, vol. 112, no.~7, 2024.

\bibitem{RanasingheTWC2025}
K.~R.~R. Ranasinghe  \emph{et. al}, ``Joint channel, data, and radar parameter estimation for afdm
  systems in doubly-dispersive channels,'' \emph{IEEE Transactions on Wireless
  Communications}, vol.~24, no.~2, 2025.

\bibitem{Svante2007}
J.~Du and S.~Signell, ``Comparison of {CP-OFDM and OFDM/OQAM} in doubly
  dispersive channels,'' in \emph{Future Generation Communication and
  Networking (FGCN 2007)}, vol.~2, 2007.

\bibitem{Rou_SPM_2024}
H.~S. Rou  \emph{et. al}, ``From orthogonal time–frequency space to affine
  frequency-division multiplexing: A comparative study of next-generation
  waveforms for integrated sensing and communications in doubly dispersive
  channels,'' \emph{IEEE Signal Process. Mag.}, vol.~41, no.~5,
  2024.

\bibitem{bemani2023}
A.~Bemani, N.~Ksairi, and M.~Kountouris, ``Affine frequency division
  multiplexing for next generation wireless communications,'' \emph{IEEE
  Transactions on Wireless Communications}, vol.~22, no.~11,
  2023.

\bibitem{tao2024daft}
Y.~Tao  \emph{et. al}, ``{DAFT}-spread affine
  frequency division multiple access for downlink transmission,'' in
  \emph{GLOBECOM 2024 - 2024 IEEE Global Communications Conference}, 2024.

\bibitem{omar2020spectrum}
M.~S. Omar and X.~Ma, ``Spectrum design for orthogonal chirp division
  multiplexing transmissions,'' \emph{IEEE Wireless Communications Letters},
  vol.~9, no.~11, 2020.

\bibitem{savaux2024}
V.~Savaux, ``Special cases of {DFT}-based modulation and demodulation for
  affine frequency division multiplexing,'' \emph{IEEE Transactions on
  Communications}, vol.~72, no.~12, 2024.

\bibitem{zakaria2012novel}
R.~Zakaria and D.~Le~Ruyet, ``A novel filter-bank multicarrier scheme to
  mitigate the intrinsic interference: Application to mimo systems,''
  \emph{IEEE Transactions on Wireless Communications}, vol.~11, no.~3, 2012.

\bibitem{nissel2018pruned}
R.~Nissel and M.~Rupp, ``Pruned {DFT-spread FBMC}: Low {PAPR}, low latency,
  high spectral efficiency,'' \emph{IEEE Transactions on Communications},
  vol.~66, no.~10, 2018.

\bibitem{pereira2021novel}
R.~P. Junior, C.~A. F.~d. Rocha, B.~S. Chang, and D.~Le~Ruyet, ``A novel {DFT}
  precoded filter bank system with iterative equalization,'' \emph{IEEE
  Wireless Communications Letters}, vol.~10, no.~3, 2021.

\bibitem{pereira2022generalized}
------, ``A generalized {DFT} precoded filter bank system,'' \emph{IEEE
  Wireless Communications Letters}, vol.~11, no.~6, 2022.

\bibitem{pereira2023two}
------, ``A
  two-dimensional {FFT} precoded filter bank scheme,'' \emph{IEEE Transactions
  on Wireless Communications}, vol.~22, no.~11, 2023.

\end{thebibliography}


\end{document}